# Deep learning networks for selection of measurement pixels in multi-temporal SAR interferometric processing

Ashutosh Tiwari *(*corresponding author*)

Civil Engineering Department

Indian Institute of Technology Kanpur,

Kanpur-208016, UP, India

Phone-+91-8090481541

Email-ashutoshtiwari796@gmail.com

Avadh Bihari Narayan*

Civil Engineering Department

Indian Institute of Technology Kanpur,

Kanpur-208016, UP, India

Phone- +91-7052325650

Email-avadhbihari096@gmail.com

Onkar Dikshit

Civil Engineering Department

Indian Institute of Technology Kanpur,

Kanpur-208016, UP, India

Phone-+91-9450346572

Email-onkar@iitk.ac.in

**\***Both authors contributed equally to this manuscript

# Deep learning networks for selection of measurement pixels in multi-temporal SAR interferometric processing

**Abstract**

In multi-temporal SAR interferometry (MT-InSAR), persistent scatterer (PS) pixels are used to estimate geophysical parameters, essentially deformation. Conventionally, PS pixels are selected on the basis of the estimated noise present in the spatially uncorrelated phase component along with look-angle error in a temporal interferometric stack. In this study, two deep learning architectures, namely convolutional neural network for interferometric semantic segmentation (CNN-ISS) and convolutional long short term memory network for interferometric semantic segmentation (CLSTM-ISS), based on learning spatial and spatio-temporal behaviour respectively, were proposed for selection of PS pixels. These networks were trained to relate the interferometric phase history to its classification into phase stable (PS) and phase unstable (non-PS) measurement pixels using ~10,000 real world interferometric images of different study sites containing man-made objects, forests, vegetation, uncropped land, water bodies, and areas affected by lengthening, foreshortening, layover and shadowing. The networks were trained using training labels obtained from the Stanford method for Persistent Scatterer Interferometry (StaMPS) algorithm. However, pixel selection results, when compared to a combination of R-index and a classified image of the test dataset, reveal that CLSTM-ISS estimates improved the classification of PS and non-PS pixels compared to those of StaMPS and CNN-ISS. The predicted results show that CLSTM-ISS reached an accuracy of 93.50%, higher than that of CNN-ISS (89.21%). CLSTM-ISS also improved the density of reliable PS pixels compared to StaMPS and CNN-ISS and outperformed StaMPS and other conventional MT-InSAR methods in terms of computational efficiency.

## Introduction

In recent decades, Synthetic Aperture Radar interferometry (InSAR) has been successfully applied in the measurement of earth surface deformation. In the initial phase of InSAR technique improvement, the Differential SAR interferometry (DInSAR) proved to be an effective tool for displacement measurement between interferometric pairs. The advent of multi-temporal InSAR methods led to further improvements, overcoming the limitations associated with the DInSAR technique (inability to reduce atmospheric error and decorrelation noise) through time series processing involving differential interferometric stacks. One of the widely used MT-InSAR technique is the Persistent Scatterer InSAR (PS-InSAR), which selects pixels with high coherence throughout the interferometric phase history, known as persistent scatterer (PS). These pixels are less affected by spatial and temporal decorrelation noise, and possess highly stable phase history. There are many realizations of the PS-InSAR method (Ferretti *et al.* 2000; Ferretti *et al.* 2001; Hooper *et al.* 2007). PS pixel selection is a very important step in the PS-InSAR processing chain. The first realisation of PS-InSAR algorithm uses the amplitude dispersion ($D_A$) to select PS pixels, under the assumption that $D_A$ can be used as a surrogate of phase stability. Further, the algorithm used an apriori temporal model to validate the selection. The performance of the algorithm was satisfactory in case of urban regions. However, for non-urban terrains where slope is facing opposite to the satellite (highly probable location for presence of PS pixel), this algorithm failed to detect PS pixels. Another algorithm proposed by Hooper *et. al.* (2007) known as the Stanford Method for Persistent Scatterer (StaMPS), employs a phase stability based approach, which overcame the limitation of $D_A$ based PS selection approach. StaMPS, instead of using apriori information about deformation, uses the spatially correlated nature of deformation. Further, PS pixels are selected based on the estimated spatio-temporally uncorrelated phase noise. Although StaMPS algorithm overcame the limitations of the previous algorithms, its PS selection process is not able to identify all possible PS pixels in both urban and non-urban terrains. In the chain of developing PS selection methods, an amplitude statistics based PS selection criteria was suggested by Ferretti et al (2011). This approach laid emphasis on the fact that PS pixels have a statistically inhomogeneous behaviour of amplitude history among its neighbourhood and a threshold on the number of statistically homogeneous pixels (SHP) was used to select PS pixels. However, the SHP based approach failed to identify PS pixels with low reflectivity (non-urban terrains where slope is facing opposite to the satellite). Out of the above mentioned algorithms, StaMPS is an open access algorithm while the other two methods are proprietary. Moreover, even with the application of the advanced MT-InSAR algorithms, more than 95% of the pixels in a series of interferograms are not selected as PS pixel, and there is scarcity of PS pixels, specifically in non-urban terrains.

Recently, various deep learning (DL) architectures have been introduced and effectively used in classification problems involving image classification and segmentation, time series prediction *etc* in the domain of computer vision (Lotter *et al.* 2017; Hori *et al.* 2017; Ren *et al.* 2018; Hasasneh *et al.* 2018). These architectures, capable of identifying hidden relationships among the system input and output parameters, are now being applied to other fields as well. In remote sensing, classification of aerial and satellite images having varied nature of reflectance, *e.g.* multi-spectral, hyper-spectral and optical images, Lidar and drone point clouds, SAR images *etc*. is now being tested using one or more, of the above mentioned architectures (Li *et al.* 2019; Kumar *et al.* 2018). Classification results using deep learning, however, in remote sensing, suffers a little due to presence of higher amount of noise, coarser spatial resolution, complex object geometry, less number of samples with very large and irregular size, compared to those used in computer vision. Hence a lot of hyper-parameter tuning is required to make the DL algorithms learn the remote sensing data with better accuracy. Some promising results are available on the classification of SAR images, which suggest implementation of these architectures on more datasets (Li *et al.* 2019; Kumar *et al.* 2019; Hu *et al.* 2019).

The selection of PS pixels takes most of the computational time, which generally requires processing millions of pixels present in a series of interferograms and requires iterations to check the convergence rate of phase standard deviation (Hooper *et al.* 2007). The time required may vary from hours to days, depending on the number of candidate pixels and the size of the dataset. Nevertheless, with near real-time deformation monitoring applications which require PS-InSAR processing at regular intervals of time, the requirement of such a large amount of processing time is cumbersome. Deep neural networks, once trained, require very little time for predicting the output, which can be a solution to large time requirements while processing large stacks of differential interferograms on a regular basis. The deep networks, can speed up the PS selection process, helping in near real time MT-InSAR processing, helping in better planning and rescue operations.

In this study, two DL based architectures are proposed for PS pixel selection in a time series of differential interferograms. Two deep networks, namely convolutional neural network for interferometric stack semantic segmentation (CNN-ISS), and convolutional long short term memory networks for interferometric stack semantic segmentation (CLSTM-ISS), are trained on different datasets of urban and non-urban terrains with varying topography, deformation characteristics, satellite geometry, and atmospheric conditions. The test dataset is an unseen real world dataset, having some characteristics similar to those of the sites used for training, so that the scalability of the models can be evaluated. A sequential model is developed, with a time series of interferometric phase images being input to the network, and pixel wise labels (PS and non-PS) as output labels for learning.

The paper is organized as follows. The next section provides insights about the conventional method of PS selection used to generate the training dataset. Later sections include description about the preparation of training datasets, the proposed networks, followed by the results and conclusion sections.

## 2. Conventional method of PS pixel selection

The primary purpose of PS-InSAR processing is to select highly coherent pixels (*i.e.* PS pixels), which are capable of providing accurate estimates of the geophysical parameters such as DEM error, deformation, atmospheric error, etc. Initially, a stack of differential interferograms are generated, which involves selection of a master image, co-registration of the slave images with respect to the master, and pair-wise cross-multiplication of the master with the complex conjugate of the slave images. Further, an earth flattening and topography component is estimated and removed from the interferograms using the master image orbit and acquisition information and an external DEM (Hooper *et al.* 2007). The wrapped interferometric phase $\phi_{ifg}^{i,x}$ for $x^{th}$ candidate pixel in $i^{th}$ resultant interferogram is shown in equation (1), where $W(\ )$, $B_\perp$, $K$ indicate wrapped phase, perpendicular baseline and a proportionality constant respectively. The resultant interferograms contain wrapped deformation phase component $\phi_{defo}$, along with phase components corresponding to orbit error $\phi_{orb}$, atmospheric effect $\phi_{atm}$, topographic residual $\phi_{topo}$ and noise $\phi_{noise}$. The phase component $\phi_{topo}$ corresponds to the leftover unaccounted error in the DEM. This error is of two types: (i) spatially correlated look angle (SCLA) error $\phi_{scla}$, which occurs mainly due to systematic errors of DEM, denoted by $\Delta h_{SC}^{x}$ and (ii) spatially uncorrelated look angle (SULA) error, $\phi_{sula}$, corresponding to spatially uncorrelated error in a DEM, represented by $\Delta h_{SU}^{x}$, shown in equation (1).

$$W\left(\phi_{ifg}^{i,x}\right) = W\left( \phi_{defo}^{i,x} + \phi_{atm}^{i,x} + \phi_{orb}^{i,x} + \underbrace{K^x B_\perp^{i,x} \Delta h_{SC}^{x}}_{\phi_{scla}^{i,x}} + \underbrace{K^x B_\perp^{i,x} \Delta h_{SC}^{x}}_{\phi_{sula}^{i,x}} + \phi_{noise}^{i,x} \right) \qquad (1)$$

with $\phi_{SC}^{i,x}$ and $\phi_{SU}^{i,x}$ bracketing the respective terms above.

Figure 1 shows the PS selection workflow. The steps involved in PS selection are described below:

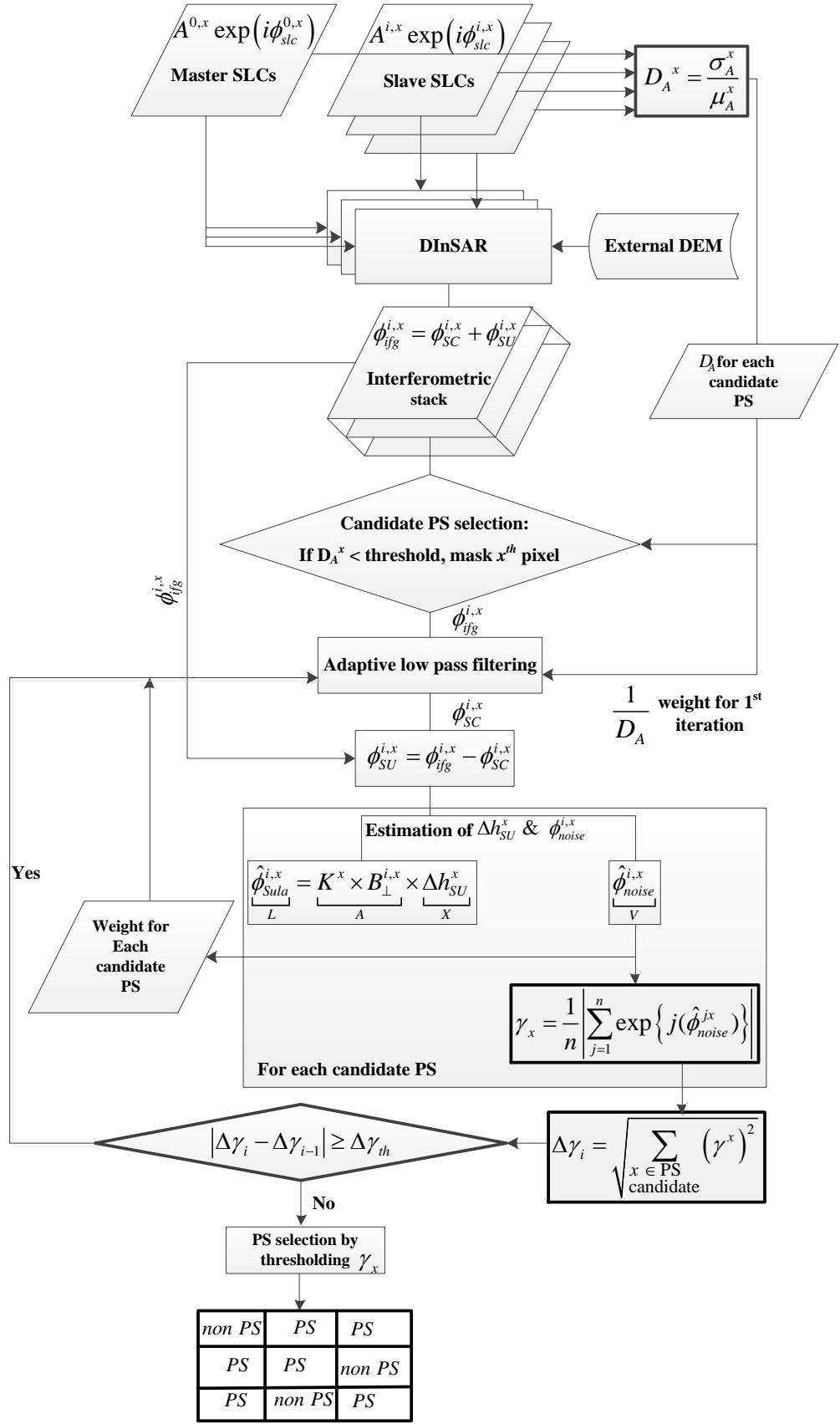

Figure 1: Workflow of PS selection by PS-InSAR approach.

*Candidate PS selection by using amplitude statistics:* Generally, for urban areas, candidate PS pixels are selected using the amplitude dispersion index $D_A$, which acts as a surrogate of the phase standard deviation due to noise $\sigma_{\phi_n}$ (Ferretti *et al.*2001; Hooper *et al.* 2007). A pixel with small $D_A$ value contains smaller $\sigma_{\phi_n}$, *i.e.* pixels with small $D_A$ are more likely to be PS pixels. However, in case of non-urban and hilly regions, even for very small values of $D_A$ (< 0.25), high phase stability (low value of $\sigma_{\phi_n}$) is not necessary. Similarly, for high $D_A$, there is a possibility of high phase stability. Consequently, it is observed that the $D_A$ based PS selection criteria works well in urban region. However, it fails in non-urban and hilly regions.

*PS selection by phase stability analysis:* A phase-based pixel selection criteria was introduced by Hooper *et al.* (2007), which improved the density of PS pixels in non-urban regions, utilizing presence of rocks, trunks of trees, and the slopes under low geometric decorrelation. PS identification is carried out by time series analysis of the phase values, subsequent to selection of candidate pixels. After removal of the flat earth and topographic phase component from the interferometric phase, some SC phase components $\phi_{defo}^{i,x}$, $\phi_{atm}^{i,x}$, $\phi_{orbit}^{i,x}$, and $\phi_{sula}^{i,x}$) and some SU phase components ($\phi_{sula}^{i,x}$ and $\phi_{noise}^{i,x}$) remain. PS selection is made by the estimation of $\phi_{noise}^{i,x}$ for every pixel in all the interferograms, and pixels with less value of $\phi_{noise}^{i,x}$ are considered to be PS pixels. To estimate $\phi_{noise}^{i,x}$, the SU phase component, $\phi_{SU}^{i,x}$, needs to be separated from $\phi_{ifg}^{i,x}$ (the complete interferometric phase). A spatial adaptive low pass filtering (ALPF) is used to separate $\phi_{SU}^{i,x}$ from $\phi_{ifg}^{i,x}$. Since spatial filtering involves phase information of the neighbouring pixels, it is vital to assign a weight to each neighbouring pixel on the basis of the quality of phase information. Thus in the first iteration, to estimate $\hat{\phi}_{SU}^{i,x}$ (the symbol ˆ, indicates estimate of the parameter), the neighbours having low $D_A$ values are given more weight (since weight is inverse of $D_A$). An estimate of the SU phase component $\phi_{SU}^{i,x}$ is computed using equation (2).

$$W\left(\hat{\phi}_{SU}^{i,x}\right) = W\left(\phi_{ifg}^{i,x}\right) - \underbrace{W\left(ALPF\left(\phi_{ifg}^{i,x}\right)\right)}_{W\left(\hat{\phi}_{SC}^{i,x}\right)} \qquad (2)$$

Perpendicular baseline $B_\perp^{i,x}$ is computed for each candidate pixel in all the interferograms using orbit information. The term $K^x \Delta h_{SU}^x$, is treated as an unknown parameter, whereas $\phi_{noise}^{i,x}$ is considered as phase residual. Equation (3) shows a nonlinear system of equations (due to wrapped phase), which is solved by initially performing a rough search of parameter space ($\Delta h_{SU}^x$), followed by a linear inversion to estimate the best-fitting model (Hooper *et al.,* 2007). Alternatively, the least-squares ambiguity decorrelation (LAMBDA) method developed by Teunissen, (1995) is also used, followed by equation (4), where $N^{i,x}$ is ambiguous $2\pi$ phase cycle in each estimated $\phi_{SU}^{i,x}$ (Kampes, 2005).

$$W\left(\hat{\phi}_{SU}^{i,x}\right) = W\left(\underbrace{K^x B_\perp^{i,x} \Delta h_{SU}^{i,x}}_{\phi_{sula}^{i,x}} + \phi_{noise}^{i,x}\right) \qquad (3)$$

$$\hat{\phi}_{SU}^{i,x} = K^x B_\perp^{i,x} \Delta h_{SU}^{i,x} + \phi_{noise}^{i,x} + 2\pi N^{i,x} \tag{4}$$

As $\hat{\phi}_{noise}^{i,x}$ is estimated, it is used for SNR estimation $(SNR^x)$, which is further used as a weight $(W^x)$ for the next iteration. The SNR can be estimated by following equation (5), where $a^{i,x}$ is amplitude of $i^{th}$ slave SLC and $\hat{\phi}_{noise}^{i,x}$ is the estimated noise of the interferograms corresponding to the $i^{th}$ slave SLC.

$$W^x = SNR^x \approx \frac{\frac{\sum_{i=1}^{n} a^{i,x} \cos(\hat{\phi}_{noise}^{i,x})}{n}}{\frac{1}{2}\left[\frac{\sum_{i=1}^{n}(a^{i,x})^2}{n} - \left(\frac{\sum_{i=1}^{n} a^{i,x} \cos(\hat{\phi}_{noise}^{i,x})}{n}\right)^2\right]} \tag{5}$$

As $\hat{\phi}_{noise}^{i,x}$ is a wrapped phase, the minimization of the total sum of square of $\hat{\phi}_{noise}^{i,x}$ ( or the absolute value of $\hat{\phi}_{noise}^{i,x}$ ) cannot be used as a criterion for optimal parameter estimation. Therefore, the coherence $\gamma^x$ (equation 6) is used as a measure of the variation in phase time series due to $\hat{\phi}_{noise}^{i,x}$. For a larger value of $\gamma^x$, variation in time series due to $\hat{\phi}_{noise}^{i,x}$ will be smaller. The unknown parameters ($\Delta h_{SU}^x$ and $\hat{\phi}_{noise}^x$) are estimated iteratively until the total sum of the coherence of all the candidate PS pixels $\left(\gamma_{total} = \sum_{x \in C_{PS}} \gamma^x\right)$ gets maximized. During an iteration, if the root mean square (RMS) change in $\gamma_{total}$ falls below a decided threshold value, it is assumed that the solution has converged.

$$\gamma_x = \frac{1}{n}\left|\sum_{j=1}^{n} \exp\{j(\hat{\phi}_{noise}^{jx})\}\right| \tag{6}$$

Finally, out of all the candidates, the PS pixels are selected in a probabilistic approach on the basis of the estimated $\gamma^x$ value. The threshold value $\gamma_{th}^x$, should be chosen in such a way that the number of false positives is minimized.

### 3. Preparation of training datasets

PS-InSAR processing requires a stack of differential interferograms as input, generally prepared with the help of DInSAR technique. Hence, input to the proposed networks was chosen to be the same in order to minimize user inputs. However, since the size of each interferogram is large, containing millions of pixels, each interferometric phase image was divided into image blocks of size 100×100 each. This allowed the network to be trained with greater computational efficiency, since using the original size would require larger computing and memory usage, and would also restrict GPU usage. Further, the number of training samples were increased, which supported better network learning. Furthermore, deformation phenomena varies across different terrains. Hence, it was necessary to train the network on interferometric images pertaining to different varieties of deformation. This is evident from the fact that many deformation prone sites are affected by landslides, land subsidence and slow recurring movements while some face large magnitude earthquakes and volcanic eruptions. Therefore, in order to account for this variability, four different study sites, having different deformation characteristics were used for training. Table 1 gives a list of study sites used for training. MT-InSAR processing of three different study sites was performed to generate three interferometric stacks. The original image dimensions were large, and each dataset generated a stack of 10 interferograms. The interferograms in these stacks were then divided into image patches of size 100×100 pixels. As a whole, the training dataset contained ~10,000 images (or image patches). Further, different temporal baselines were used in the three datasets to make the DL network account for the temporal variability (Table 1). Spatial baselines for all generated interferograms were between 100 and 400 m. The training dataset was divided into two parts, (i) for training and (ii) for validation. The validation dataset was used to evaluate the model performance during training, thereby helping in hyper-parameter tuning. Apart from the training and validation datasets, interferometric images of the Kathmandu city were used as the testing dataset. Different from the training and validation datasets, the testing images were used to evaluate how the networks performs on real world unseen data, as suggested by deep learning experts (Shah, 2017; Brownlee, 2017). Once the networks were trained, evaluation was carried out on the test data, and predicted blocks of 100×100 pixels were combined to form the labelled image of the complete area under evaluation. Figure 2 shows the training, validation and test datasets. Random sampling was used to select test samples (images) from the training data, to avoid overfitting during network learning.

Table 1. Training, testing and validation datasets used for training the proposed network.

| S.No. | Site | Original image dimension ($R \times Az$) | #images (100×100) | Temporal coverage | Nature |
|---|---|---|---|---|---|
| 1 | New Delhi | 3500×10500 | 3675 | 15 Nov 2015 to 12 Sept 2018 | Urban with flat terrain |
| 2 | Ahmedabad | 2200×7700 | 2500 | 27 Sept 2016 to 06 Feb 2017 | Urban |
| 3 | Nainital | 5000×8000 | 4000 | 4 Apr 2018 to 5 May 2019 | Non-urban with hilly terrain |
| 4[†] | Kathmandu | 700×2900 | 203 | 24 Mar 2015 to 7 Nov 2015 | Semi-urban with hills and flat areas |

'†' indicates test dataset, $R$ and $Az$ denote range and Azimuth respectively.

Since PS-InSAR based PS pixel selection has several realizations with algorithms proposed by Ferretti *et al*. 2001, Kampes 2006; Hooper et al 2007; Agram 2010; Ferretti et al. 2011, any of these methods could be used for the preparation of training labels. However, the widely used StaMPS algorithm, which is an open access algorithm, was selected, so as to make the implementation of the proposed methodology easier for most of the researchers working in the field of InSAR. Further, the training labels were generated using Sentinel-1 interferometric Wide Swath (IW) images, which have global coverage and are made freely available by the European Space Agency (ESA). Currently, the ESA provides its own software integrated with the StaMPS method for PS-InSAR processing. The above factors motivated to select the StaMPS based PS selection results for generating training labels. Nevertheless, proprietary datasets including higher resolution SAR images and corresponding software can also be used with the proposed networks.

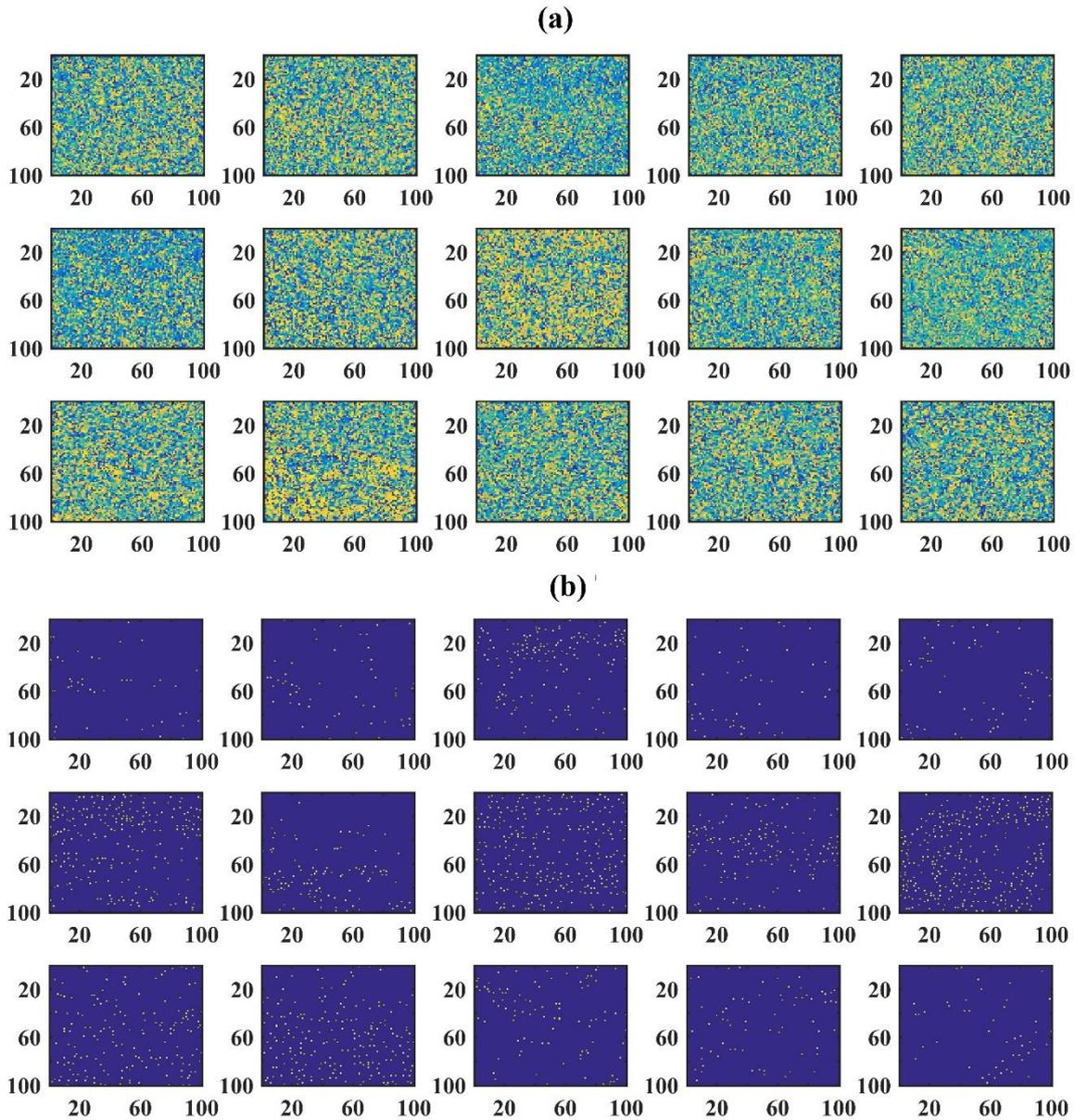

Figure 2. Example images of the Training dataset used during network training. (a) Wrapped interferograms (b) Labelled images with yellow dots representing the PS pixels

A major challenge in pixel-wise classification of interferometric images was class imbalance, *i.e.* more than 95% of the pixels in a series of interferograms belong to one class (non-PS pixels). Numerous solutions to the problem of class imbalance have been investigated. Some of the solutions involve (i) giving higher weights to the samples of minority class, (ii) assigning class ratios while training, (ii) synthetically oversampling the minority class samples to bring class balance (iv) under sampling the majority class (v) improving the training dataset, (vi) using different loss functions during training to penalize false positives and true negatives, thereby trying to avoid overfitting, etc. (Data Science Stack Exchange, 2016; Stack Overflow, 2017; Sabinasz, 2019; Johnson and Khoshgoftaar, 2019). Suitability of one of these solutions or a combination of these totally depends on the problem at hand and the trade-offs associated with the problem statement.

## 4. Architecture of the proposed networks

The traditional problem statement of PS pixel selection was formulated into a semantic segmentation task (also known as pixel wise classification) by the proposed deep learning networks. Further, contrary to the most commonly developed and deployed object classification based DL methods, which focus on learning spatial patterns in the input samples, the proposed networks focussed on learning the relation between the absolute interferometric phase values and their temporal coherence.

### 4.1 *CNN-ISS architecture*

Figure 3 shows the CNN-ISS network architecture. It contains an input layer, four convolutional (conv) layers (each followed by a batch normalization (BN) layer), a dropout layer and a fully connected (FC) layer as the final layer for pixel-wise classification (or segmentation). The output is a labelled image corresponding to the stack of interferograms, with each pixel showing a semantic label. The label is either PS or Non-PS, similar to the problem of binary segmentation. The generalized mathematical representation of the network is shown in equation (7), where $f$ is the CNN-ISS network operation, $x$ is the input data, $w$ is a set of weights and biases, and $y$ is the predicted output. The same representation is applicable to the other proposed network CLSTM-ISS, with $f$ being the CLSTM-ISS operation.

$$y = f(x, w) \qquad (7)$$

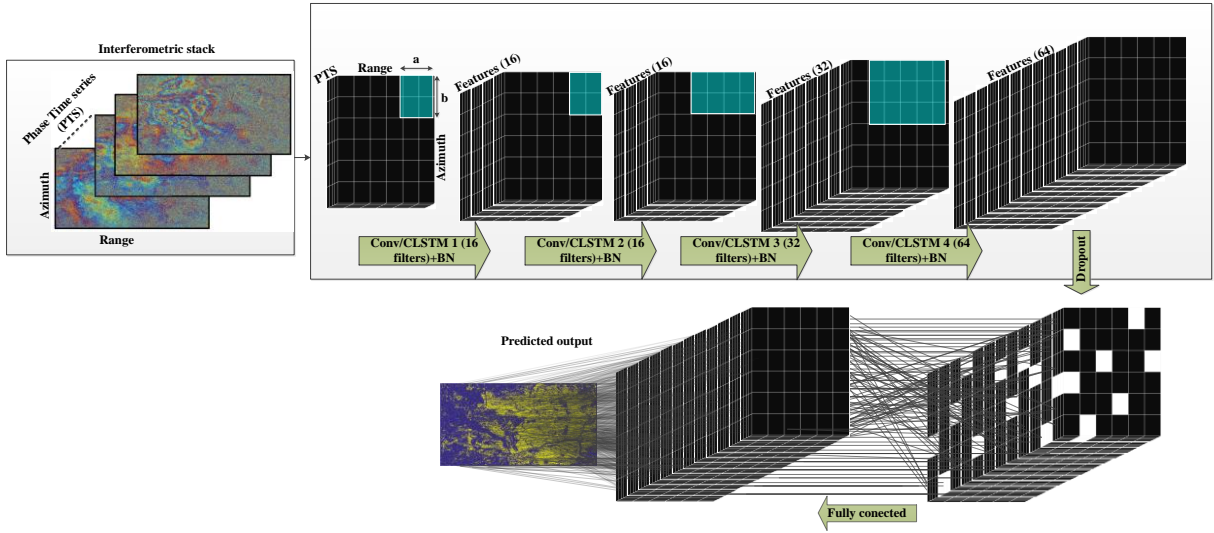

Figure 3. Architecture of the proposed networks

The input layer contains a stack of interferometric images, each of size $W \times H$ (width and eight respectively). This makes the input layer dimension as ($n, W, H, b$) with $n$ and $b$ representing the number of samples and the number of associated bands respectively. No computation is performed in this layer, and simply the input is passed to the next layer. The input layer is followed by the convolutional (*conv*) layers, whose objective is to detect feature maps from the interferometric images. The convolutional layer performs a convolution operation on the input images, as shown in equation (8), where *Im* and *k* represent the input image and kernel respectively, $k_1$ and $k_2$ denote the filter dimensions, '\*' represents the convolution operation and *F* denotes the obtained feature map (Skalski, 2019).

$$F[w,h] = (Im*k)[w,h] = \sum_{m=1}^{k_1}\sum_{n=1}^{k_2} k[m,n] Im[w-m, h-n] \qquad (8)$$

The Rectified Linear Unit (ReLU) function is used with every *conv* layer for activation, converting all negative values to zero while keeping the positive values as it is (equation 9). The transformed output (after activation) is sent to neurons of the next layer (Udofia, 2018).

$$ReLU(x) = max(0, x) \qquad (9)$$

A BN layer is added after every *conv* layer, which normalizes the output of the previous activation layer by subtracting the batch mean and dividing by the batch standard deviation. These layers are generally used to increase the stability of the network. An element wise dropout layer is added between the last conv+BN and the FC layer. This layer randomly sets pixels of feature maps as zeros, which makes the proposed network robust to noisy interferometric phase, which is present in major portion of the interferograms. Unlike other networks, the dropout layer in both CNN-ISS and CLSTM-ISS is applied before the FC layer, bringing noise augmentation in the network, and making the FC layer learn in the presence of noise. The FC layer is the last layer of the network, which results in a pixel-wise label of the output image. Each unit (neuron) in the FC layer is connected to every other unit of the previous layer. For this layer, the 'sigmoid' activation function is used, whose operation is shown in equation (10). Finally, the labelled image contains class probabilities for each pixel, giving a segmented image. Configuration details of the CNN-ISS architecture is shown in Table 2.

$$\sigma(x) = \frac{1}{1+e^{-x}} \qquad (10)$$

Table 2. CNN-ISS network configuration. '#' indicates number, and $N$ represents the number of training samples

| Layer (type) | #filters | Output dimension |
| --- | --- | --- |
| Input | | $N,100,100,10,1$ |
| (*conv*+BN)$_1$+relu | 16 | $N, 100, 100, 10, 16$ |
| (*conv* +BN)$_2$+relu | 16 | $N, 100, 100, 10, 16$ |
| (*conv* +BN)$_3$+relu | 32 | $N, 100, 100, 10, 32$ |
| (*conv* +BN)$_4$+relu | 32 | $N, 100, 100, 10, 64$ |
| Dropout | | $N, 100, 100, 10, 64$ |
| FC | | $N, 100, 100, 1$ |

### 4.2. CLSTM-ISS architecture

The second type of network developed in this study involved the use of spatio-temporal input to classify PS pixels. This is accomplished using a combination of the convolution and LSTM (*convlstm*) layers. Individual LSTM layers are capable of finding time series relationships between input and predicted parameter(s) by controlling the flow of temporal information using gates. In addition to the four layers inside an LSTM cell (three of which are sigmoid and one is a hyperbolic tangent (tanh) layer), if a convolution layer is added, a *convlstm* cell is obtained. In *convlstm*, the convolution operation substitutes matrix multiplication for all gates, helping in detection of spatial features. The *convlstm* layer uses the current time step input $X_t$ and the previous state of local neighbours $S_{t-1}$ to find the output state $Y_t$, as shown in Figure 4. The horizontal line at the top of the network represents the cell state $S$ through which the temporal information is carried. Operations of the *convlstm* layer are shown in equation (11), where $(X_1, X_2,...X_t)$, $(S_1, S_2,...,S_t)$, and $(\hat{Y}_1, \hat{Y}_2,...,\hat{Y}_t)$ represent input samples, cell state and hidden state outputs respectively, and $i_t, f_t, o_t$ denote input, forget and output gates respectively. The symbol '∗' denotes the convolution operator, '∘' represents the Hadamard product and $W$ denotes the weight matrix.

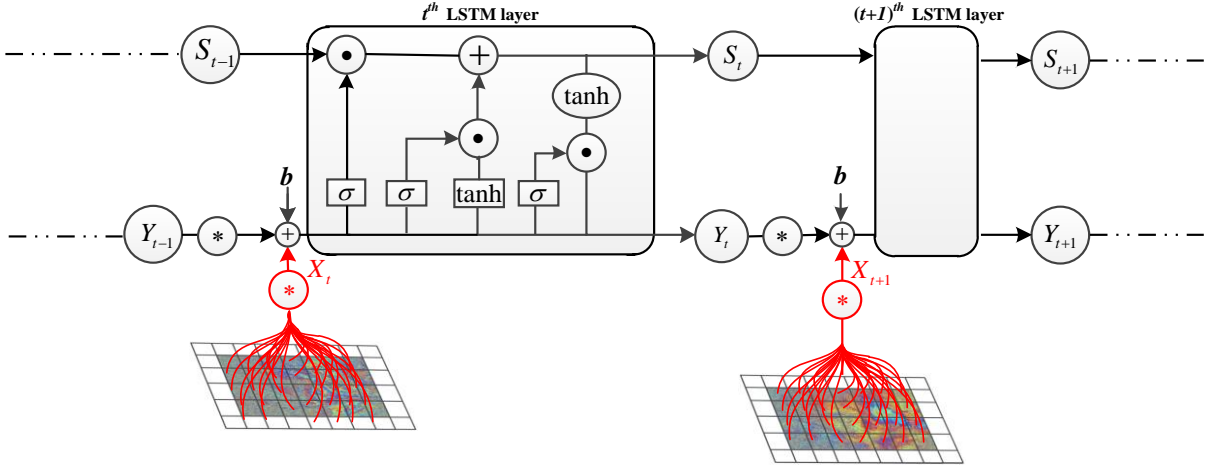

Figure 4. Basic structure of a *convlstm* cell. The convolution operation is shown by '∗' symbol.

$$\begin{aligned}
i_t &= \sigma\left(W_{xi} * X_t + W_{yi} * \hat{Y}_{t-1} + W_{si} * S_{t-1} + b_i\right) \\
f_t &= \sigma\left(W_{xf} * X_t + W_{yf} * \hat{Y}_{t-1} + W_{sf} * S_{t-1} + b_f\right) \\
S_t &= f_t \circ S_{t-1} + i_t \circ \tanh\left(W_{xs} * X_t + W_{ys} * \hat{Y}_{t-1} + b_s\right) \\
o_t &= \sigma\left(W_{xo} * X_t + W_{yo} * \hat{Y}_{t-1} + W_{so} * S_t + b_o\right) \\
\hat{Y}_t &= o_t \circ \tanh(S_t)
\end{aligned} \qquad (11)$$

The CLSTM-ISS architecture can be derived from the CNN-ISS architecture shown in Figure 3, with *conv* layers replaced by *convlstm* layers. This architecture was trained on a stack of 10 interferograms, providing 10 time steps to the problem of spatio-temporal prediction. The network contains an input layer, two *convlstm* layers (each followed by a BN layer), one *conv* layer, a dropout, an FC and an output layer. The input layer contains a time series of interferometric images with $t$ time steps, each of size $w \times h$, where $w$ and $h$ denote the image width and height respectively. This makes the input layer dimension as ($n, t, W, H, b$) with $t, n$ and $b$ representing the number of time steps, samples and channels (bands) respectively. Both the *convlstm* layers, supplemented with ReLU activation, are followed by a *conv* layer. A dropout layer is added next, followed by an FC layer. Apart from the *convlstm* layer, the remaining layers perform similar to the respective layers of CNN-ISS network. Table 3 shows the architecture dimensions. The number of time steps was kept flexible to make the network learn and predict output with different number of time steps, which in this study is the number of interferograms.

Table 3. CLSTM-ISS network configuration. '#' denotes number.

| Layer (type) | #filters | Output dimension |
|---|---|---|
| (*convlstm*+BN)$_1$+relu | 16 | #samples, #timesteps,100, 100, 16 |
| (*convlstm*+BN)$_2$+relu | 16 | #samples, #timesteps 100, 100, 16 |
| (*convlstm*+BN)$_2$+relu | 32 | #samples, #timesteps 100, 100, 32 |
| (*conv*)$_1$+relu | 32 | #samples, #timesteps, 100, 100, 64 |
| Dropout | | #samples, #timesteps, 100, 100, 64 |
| FC | | #samples, 100, 100, 1 |

Pooling layers were avoided in the proposed architecture for various reasons. These layers downsample the input image by generally taking the average/maximum of all values contained under an applied filter (*e.g.* of size 2x2). The layers bring translational invariance to the network, and reduce computational time by downsampling. However, PS pixels are not found in a cluster and do not follow any spatial structure. Rather, these pixels are selected on the basis of correlation among the neighboring pixels. The neighborhood range is decided based on the extent of SC deformation, and not on that of other SC components (*e.g*, $\phi_{atm}$, $\phi_{sula}$). A PS pixel could be surrounded by either a DS or NS pixel and it is expected that a PS pixel contains SC information in the neighborhood. However, since DS or NS pixels follow a uniform distribution, it is not mandatory that the maximum phase value in a kernel window will be that of a PS pixel. There is uniform probability that the maximum or average phase selected by pooling could be that of NS or DS pixel as well. Another implication of applying pooling is that the filter size is increased in the next convolutional layer, with the same ratio with which the feature maps were reduced. Thus, the network is imposed to learn features with phase components of larger spatial extent, *i.e.* learning on the basis of spatially correlated components other than deformation. This could increase the number of false positives. Hence, as a whole, pooling could reduce the computational time during training, but was not applied due to the chance of inaccurate learning.

**4.3. Treatment of imbalance problem**

The dataset used in this study was highly imbalanced, *i.e.* more than 95% of the pixels in a series of interferograms belong to one class (non-PS pixels). To tackle this problem, the *$f_1$-loss* was used as the loss function, which used the *$f_1$-score* maximization as the criteria minimize the loss (Ng, 2019). With two classes (PS and Non-PS), the PS class was given a weight of 200 and Non-PS class was given a weight value 1. In addition, since the *$f_1$-score* is non-differentiable, hence the calculation of difference between the true label and the predicted label was calculated probabilistically. For example, if the true label was Non-PS (or zero), and the predicted probability came out to be 0.4, accuracy was evaluated as 0.6 false positive and 0.4 true negative. Further, accuracy, precision $TP/(TP+FP)$, recall $TP/(TP+FN)$, and *$f_1$-score* $2\times(precision\times recall)/(precision+recall)$ were used as evaluation metrics during training.

Both the architectures were trained using high performance computing (HPC) facility at the Indian Institute of Technology Kanpur. Python programming environment, along with the keras library and Tensorflow library (backend) were used for writing and testing programs. CNN-ISS and CLSTM-ISS were trained on 400 and 300 epochs respectively, using the training dataset mentioned in Section 3. An early stopping criteria of 20 epochs was used, *i.e.* if for 20 continuous epochs, the validation loss did not improve, the training was stopped. Adam optimizer was used with learning rates of 0.01 and 0.001 for CNN-ISS and CLSTM-ISS respectively. Once trained for sufficient number of epochs (converged), the architecture weights and biases were saved, and were later used for producing PS pixel classified images.

## 4. Results and discussion

The complete labelled images of the training datasets generated from StaMPS based PS-InSAR processing (divided into chunks of 100×100 image blocks for training), are shown in Figure 5. These images were obtained from processing the interferometric stacks of Sentinel-1 IW images belonging to New Delhi, Nainital and Ahmedabad. The complete set of pixels in the labelled images was a combination of PS and Non-PS pixels, where each training label was labelled as either PS or non-PS pixel. While preparing training labels using these real world datasets, higher emphasis was laid on minimizing the number of false positives (actually a non-PS but labelled as a PS pixel). This was accomplished by setting a higher coherence threshold during StaMPS based PS candidate selection (equation 6). Hence, although the number of the labelled PS is less, the false-positive PS pixels were very less in number compared to the true PS pixels. Similarly, minimization of the number of false negatives (actually a PS but labelled as a non-PS pixel) in the set of the labelled non-PS pixels was also considered. However, since it was already known that in any study site, the number of non-PS pixels is very large compared to that of the PS pixels, the number of false negative non-PS pixels in the set of labelled non-PS pixels had to be very less compared to that of the true non-PS pixels. Nevertheless, with training samples generated in the above manner and the chosen architecture of the proposed networks, the difference in selection of different number of PS pixels was expected.

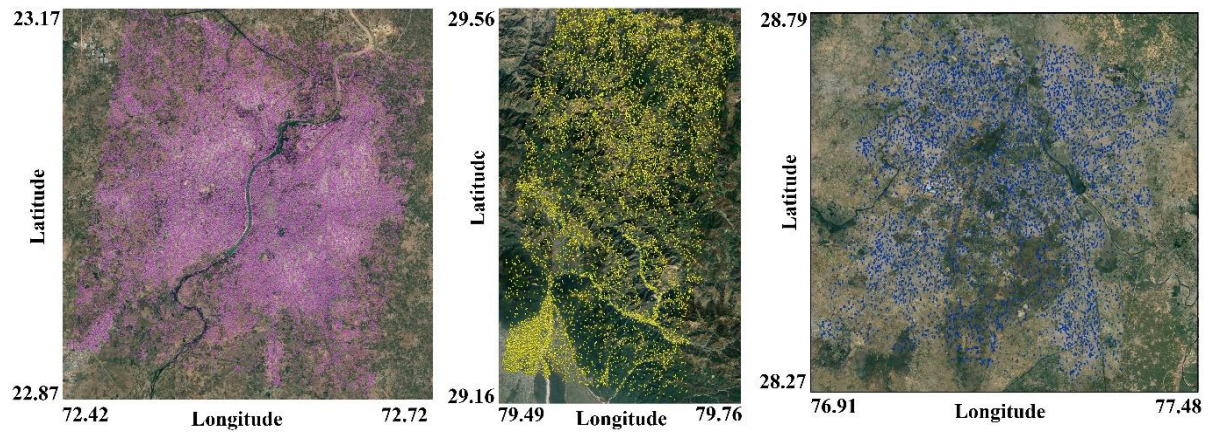

Figure 5. Complete labelled images corresponding to the study sites used for training. (a), (b) and (c) represent labelled images for New Delhi, Nainital and Ahmedabad respectively. Point objects in each image represent the PS pixels.

Figure 6 shows the PS pixel selection results of the Kathmandu city dataset, which is the unseen dataset used for testing in this study. The predicted output images, each of size 100×100 were combined to generate a complete deformation map. Figure 6 (a) shows a Google earth image of the Kathmandu city. Some areas in the image are marked with white boundary polygons, showing pixels belonging to different categories such as forest and vegetation, uncropped field, water body, etc. In addition, areas affected by foreshortening/layover (considering SAR image acquisition) are also shown. The associated R-index image is shown in Figure 6 (b), which gives an indication of where PS pixels are likely to occur. R-index values lying between 0.5 and 0.9 belong to slopes facing towards the satellite (under lengthening) and have a high probability of being a PS pixel due to low geometric decorrelation noise. R-index values other than the value range 0.5 and 0.9 belong to either the slopes opposite to the satellite LOS (affected by foreshortening or layover) or areas under shadowing. The probability of finding PS in such regions is very low. Figure 6 (c) shows a classified image of the Kathmandu city with four classes, namely man-made objects, forest and vegetation, river and uncropped land. In the urban region, where man-made objects are abundant, there is higher likelihood of finding PS pixels, due to presence of a dominant scatterer (building corners, concrete structures, etc) in resolution cells. The urban areas, uncropped land, water bodies, forest and vegetated areas are mostly devoid of dominant scatterers and are also affected by temporal decorrelation. PS pixels are thus rare in these areas.

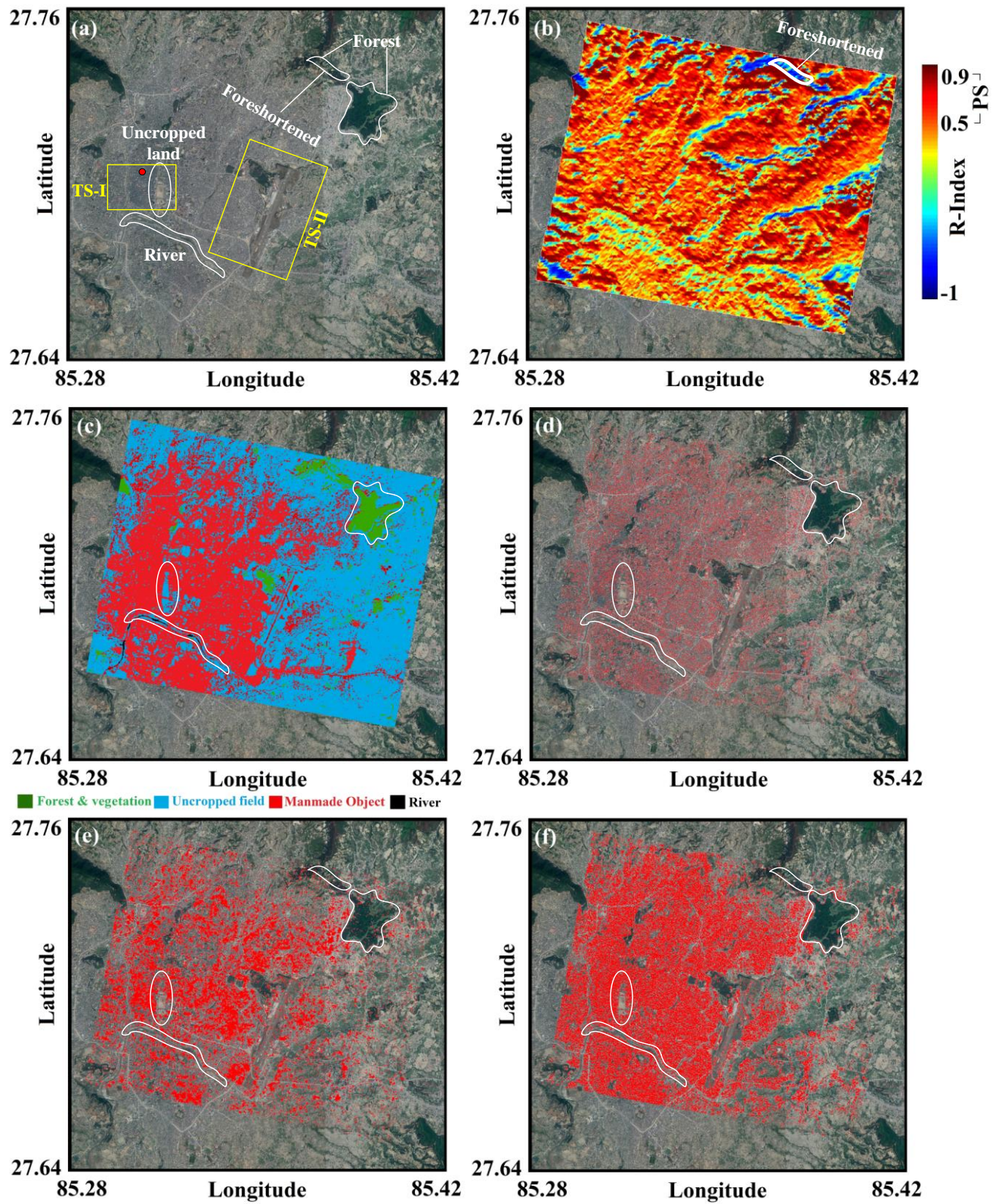

Figure 6. Results of PS selection. (a) Study area containing different features used for performance evaluation and two test sites TS-I and TS-II for detailed analysis. Red dot indicates the location used for analysis of time series velocity estimates (b) computed R-index (c) Classified map. (d), (e) and (f) show the pixel selection output from StaMPS processing, CNISS prediction and CLSTM-ISS prediction respectively.

The problem of PS selection is dissimilar from other supervised classification problems due of lack of true labels. A pixel could be a PS for an interferometric stack, and the same pixel could be non-PS for another stack generated for the same study site, but with different view angles and acquisition times. During training of the proposed networks, accuracy, *$f_1$-score*, *precision* and *recall* were used as evaluation metrics along with the *$f_1$-loss* to avoid overfitting and to make the network better distinguish the interferometric phase characteristics of the under-sampled PS pixels. However, with the lack of true labels, using these metrics for quality evaluation would provide an incorrect estimate of the classifier (network) performance. Thus, a combination of R-index and classified image was used for qualitative evaluation. R-index and classified image provided information about the geometric and temporal decorrelation respectively, to yield a probabilistic measure of the presence of PS pixel at a certain location. From previous literature, it is known that pixels belonging to urban areas (due to presence of buildings and other man-made structures) and slopes facing opposite to the satellite LOS (area under lengthening) possess higher probability of being a PS (Notti *et al.* 2011).

The above mentioned combination was used to decide the appropriate pixel labels for comparing the pixel selection results of StaMPS and of the two proposed architectures. The decision was made since the widely accepted StaMPS based PS results do not attain true labels. Hence, it was better to select the best possible estimates as the reference and compare it with all other classification results for optimal judgement. In addition, to visualize the pixel-wise classification aptitude of the proposed networks, different test sites characterized by different land cover and topography were selected for detailed analysis. StaMPS based PS selection output is shown in Figure 6 (d). From this output, it is apparent that the StaMPS method mostly detected man-made objects as PS pixels. However, the density of such points was low as compared to the density observed in the classified image (Figure 6 (c)). Further, a few PS pixels were detected in the areas belonging to 'forest and vegetation' class (marked with a white boundary polygon containing green patches in Figure 6(c)), where pixels are generally affected by temporal decorrelation. In case of foreshortened areas, the observed density was higher compared to that of forested areas. A small number of pixels were seen on the river, which may well be wrongly detected PS pixels.

Results of semantic segmentation predicted using the CNN-ISS weights and biases are shown in Figure 6 (e). The estimates of StaMPS and CNN-ISS are different, and it is noticeable that the number of PS pixels (as a whole) detected by CNN-ISS was more than that than detected by the StaMPS method. The predicted image from CNN-ISS had higher PS pixel density in man-made areas (see Figures 6 (c) and (e)), which was expected due to the presence of dominant scatterers in such regions. However, in the regions such as forest, vegetation, water body and uncropped land, a very small number of PS pixels was detected. However, for slopes facing opposite to the satellite LOS (foreshortened area shown in Figure 6 (b)), CNN-ISS successfully classified most of the pixels as non-PS. This throws light on the potential of the network in separating PS and non-PS pixels depending upon the nature of land use and land cover.

Figure 6 (f) shows the PS selection image predicted by the CLSTM-ISS architecture. Similar to the estimates of the CNN-ISS, the predicted image contained more number of PS pixels compared to that detected using the StaMPS approach. The obtained density of PS pixels in case of city is even higher than that of CNN-ISS. Further, in the non-urban terrains (vegetation, river and uncropped land) and slope facing towards the satellite LOS, CLSTM-ISS successfully identified all the non-PS pixels, which is evident from Figures 6 (b) and (c).

In order to better comprehend the pixel-wise classification aptitude of the proposed networks, two different test sites TS-I and TS-II, characterized by different land use and land cover were selected. These test sites are marked in Figure 6 (a). Figures 7 and 8 provide insights about the capability of the proposed networks by comparing the obtained estimates of the chosen test sites at higher level of detail. In the first phase of quality assessment, efficiency of the proposed networks is evaluated on buildings, playground (uncropped land) and river (Figure 7). Figure 7 (a) shows a Google earth image of TS-I, containing the above three features. Since PS pixels strongly correspond to those resolution cells which contain a dominant scatterer (*e.g.* buildings), therefore the PS pixel density was expected to be high at locations other than river and playground. It was observed that the StaMPS method correctly identified almost all the pixels as non-PS in the river and playground (Figure 7 (b)). This shows that the StaMPS estimates contained very small number of false positives in water body and uncropped land. However, in the area full of buildings, the observed PS density was very low (white points in figure 7(b)). Similar to the PS selection by StaMPS, both CNN-ISS as well as CLSTM-ISS correctly labelled almost all the pixels as non-PS in the river and uncropped land. This indicates that both the proposed networks were able to correctly label pixels in water body and uncropped land, and the results were comparable with that of that of the StaMPS method. However, in the area covering buildings, the density of PS pixels detected by both CNN-ISS and CLSTM-ISS is much higher than that obtained by StaMPS. Thus, in the city where man-made structures are plentiful, both CNN-ISS and CLSTM-ISS outperformed the StaMPS method.

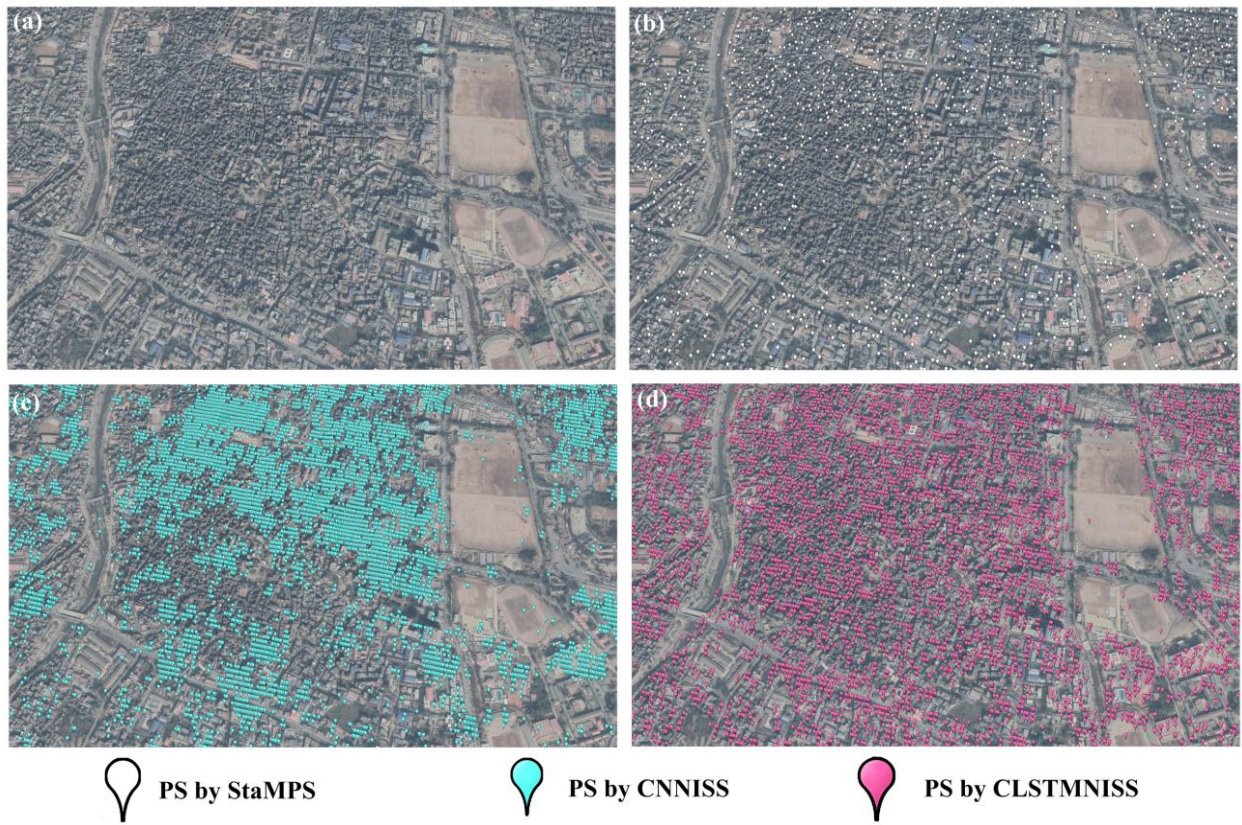

Figure 7: Detailed view of test site TS-I shown in Figure 6 (a). (a) Google earth image of test site, (b), (c) and (d) show PS selection results of StaMPS, CNN-ISS and CLSTM-ISS overlaid on Google earth image respectively.

It is critical to notice that the PS pixels found by CNN-ISS were in clusters, very dense at some locations and very sparse at some other locations. Conversely, the PS density of CLSTM-ISS was found to be uniform at most of the locations. The presence of phase noise causes reduction in both spatial and temporal correlation. Since PS selection is totally based on phase noise, assessment of both spatial and temporal correlation of interferometric phase history may be a better approach to identify PS pixels. But the learning of *conv* layers in CNN-ISS provided more weight to spatial correlation as compared to temporal correlation during training. As a result, PS pixels were detected in a bunch on the basis of spatially correlated phase component only. Further, the lack of participation of temporal correlation in the estimation of weights and biases resulted in a non-uniformly distributed set of PS pixels in case of city. On the other hand, CLSTM-ISS used a combination of *conv* and LSTM layers (*convlstm*), utilizing both spatial and temporal correlation to optimize the weights and biases of the network. Consequently, the detected PS pixels selected were uniformly distributed. Since the CNN-ISS obtained very high density of PS pixels at some places in the city, it can also be argued that CNN-ISS had poor density in some regions but at some places, the performance is compensated with higher density, but this statement does not hold valid. The back scattered signal from a scatterer follows the *sinc* impulse response function in range and azimuth, and the dominant scatterers in PS can dominate the response of the adjacent pixel also. Therefore, many researchers suggest that it is desirable to select only the pixel having higher temporal coherence out of the set of pixels dominated by the same dominant scatterer (Hooper *et al.* 2007; Agram and Zebker, 2007). It is clearly observed from Figure 7 (c) that most of the selected PS pixels were in clusters, which may well be selected because of the dominant scattering of adjacent PS pixel(s). However, in case of CLSTM-ISS, the density of PS pixels was high and well distributed, unlike in CNN-ISS. This proved the superior performance of CLSTM-ISS in urban areas. In addition, by comparing Figures 7(b), (c) and (d), it is inferred that the CLSTM-ISS detected less false negatives as compared to StaMPS and less false positives compared to CNN-ISS in urban regions. One possible reason of the improved performance of CLSTM-ISS over CNN-ISS is the better learning capability of *convlstm* layers used in the CLSTM-ISS over the *conv* layers in CNN-ISS.

Another test site (TS-II) shown in Figure 8 (a), covering the forest and places with greenery (vegetation), was used to evaluate the performance of the proposed networks in non-urban areas. Figures 8 (b), (c) and (d) show the PS pixels selected by StaMPS, CNN-ISS and CLSTM-ISS in the area around Tribhuvan International Airport, Kathmandu. Since in forest and vegetated areas, the temporal decorrelation is very high, the probability of finding PS pixels is close to zero. In figure 8 (a), all the green areas are either forest or vegetation. Both StaMPS and CNN-ISS detected some PS pixels (false positive) in the forest and vegetated area. However, CLSTM-ISS detected very less number of PS pixels in such areas. This shows the superiority of the CLSTM-ISS over StaMPS and CNN-ISS in reducing the false positives in case of forests and vegetation. In Figure 8 (a), within the forest area (shown with white polygon) there are a few buildings (marked with circles). These buildings are correctly classified as PS by CLSTM-ISS, which shows its efficacy to detect PS in adverse conditions (i.e. when PS is surrounded by noisy pixels).

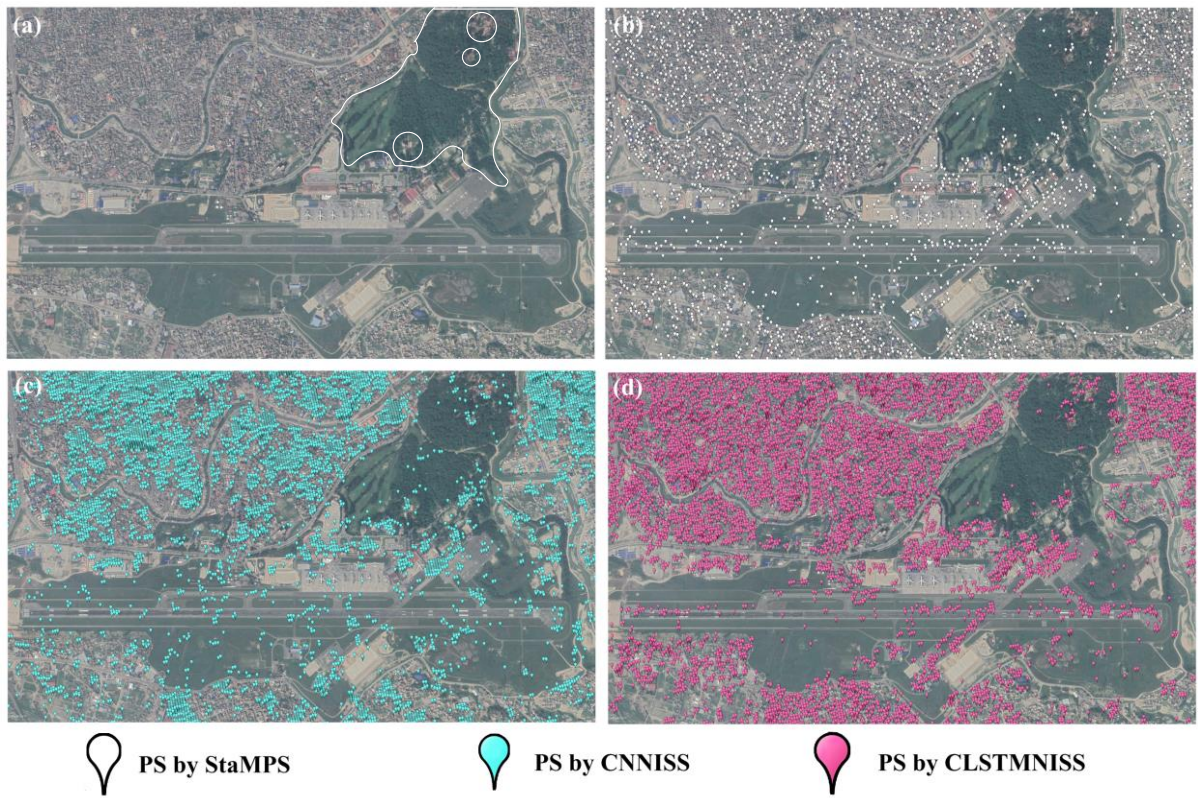

Figure 8. Detailed view of test site TS-II shown in Figure 6 (a). (a) Google earth image of test site, (b), (c) and (d) show PS selection results of StaMPS, CNN-ISS and CLSTM-ISS overlaid on Google earth image respectively.

Table 4 shows an area-wise comparison of the percentage of PS pixels obtained for StaMPS, CNN-ISS and CLSTM-ISS methods. It is observed that the density of PS pixels in areas containing man-made objects, *i.e.* city was the highest for CLSTM-ISS (52.9%). In addition, in lengthened areas where more number of PS pixels are likely to be PS, CLSTM-ISS obtained the highest percentage (48.1%) of PS pixels. The percentage in uncropped land, river and foreshortening/layover/shadowing affected areas (areas mostly containing noisy scatterers) was comparable to the other two methods. In case of forest and vegetation, CLSTM-ISS obtained the least percentage of PS pixels (3.4 %). For CNN-ISS, the percentage of PS pixels was greater than that of StaMPS for areas containing man-made features. Further, CNN-ISS correctly separated noisy scatterers in forests, river and uncropped land and is comparable to the StaMPS method. From the above area-wise comparison of PS selection, it can be stated that CLSTM-ISS outperformed the other two methods in terms of PS pixel selection and separation of classes.

Table 4. Analysis of area-wise detection of PS pixels by StaMPS, CNN-ISS and CLSTM-ISS

| S.No | Scatterer | Ground characterization | % of detected PS pixels | | |
|---|---|---|---|---|---|
| | | | StaMPS | CNN-ISS | CLSTM-ISS |
| 1 | PS | City | 16.7 | 43.2 | 52.9 |
| 2 | | Lengthening | 19.5 | 27.3 | 48.1 |
| 3 | | Forest and Vegetation | 7.1 | 13.5 | 3.4 |

| 4 | Non-PS | Uncropped land | 3.7 | 4.3 | 1.6 |
| 5 | | River | 2.1 | 2.5 | 2.2 |
| 6 | | Foreshortening/Layover/shadowing | 3.4 | 6.2 | 3.6 |

Table 5 shows the number of PS pixels detected by StaMPS, CNN-ISS and CLSTM-ISS, and the corresponding computational time required. The order of increase in the number of PS pixels is StaMPS → CNN-ISS → CLSTM-ISS. The validation accuracy is high for both CNN-ISS and CLSTM-ISS (89.21 and 93.50 percent respectively). In case of CLSTM-ISS, there is an improvement in both the validation accuracy and the validation loss.

Table 5. Analysis of PS pixel extraction and computational time requirements

| S. No | Method | # PS pixels | Computational time (minutes) | Training accuracy | Validationa ccuracy | Validation Loss |
|---|---|---|---|---|---|---|
| 1 | StaMPS | 38286 | 108.2 | | | |
| 2 | CNN-ISS | 169241 | 5.3 | 90.13 | 89.21 | 0.72 |
| 3 | CLSTM-ISS | 192177 | 8.2 | 95.42 | 93.51 | 0.64 |

Similar to the estimates of StaMPS algorithm for the testing dataset, the estimates of the training dataset also had similar characteristics. Many of the true PS pixels could not have been labelled as PS pixel in StaMPS estimates, specifically in areas containing man-made objects or in areas under lengthening. StaMPS based pixel selection may have failed to accept these true PS pixels during the phase stability estimation. However, due to the better treatment of spatially correlated phase characteristics and the temporal phase stability in the parameter estimation, CLSTM-ISS learning was closer to the true selection of PS pixels. CNN-ISS, which did not give special attention to the temporal interferometric phase variation, found difficulty in discriminating between PS and non-PS pixels where the analysis of temporal decorrelation was critical. Nevertheless, the proposed networks, one trained, can be readily used to generate pixel selection maps with better density as obtained with the StaMPS method itself, thereby helping in the improvement of unwrapping and deformation pattern recognition.

An alternative to choosing the StaMPS based PS selection results as training data was to use the results of proprietary PS-InSAR algorithms for generating training maps with higher density of PS pixels. However, this could reduce the scalability of the developed networks, since the implementation details of these algorithms are not openly accessible, and hence not available to many researchers. Further, the proposed architectures could be trained using the results of these algorithms, and the learned weights and biases could be made available for transfer learning (Brownlee, 2019). However, this would restrict the users from modifying and improving upon the developed network architectures, which took a long time to be refined for hyper-parameters. However, with the knowledge that the proposed DL methods have improved the PS selection process, new pixel selection maps can be generated for more interferometric datasets. These labelled images (predicted using the trained weights and biases of CNN-ISS and CLSTM-ISS) can become training samples to further improve the training and prediction.

## 5. Qualitative analysis of PS selection

In order to statistically assess the quality of PS pixels detected using the proposed architectures, a reliable index, known as similar time series interferometric pixel (STIP) was employed. STIP is a measure of noise in phase time series (Narayan *et al.*, 2018a; 2018b). In case of PS pixels, the number of STIP in a neighbourhood are expected to be high compared to that of non-PS pixels. Figures 9 (a), (b) and (c) show the comparison of the number of STIP detected in the neighbourhood of PS pixels selected by StaMPS, CNN-ISS and CLSTM-ISS, respectively. It is found that 92.49% of the total PS pixels selected by the StaMPS method have more than 35 STIP (a threshold generally used to define a coherent PS pixel) in the neighbourhood defined by a window of 25×5 pixels (Table 6). For CNN-ISS, 80.10% pixels have more than the aforementioned threshold STIP in the neighbourhood, which is less that than those detected by using the StaMPS method. However, for CLSTM-ISS, 97.01% of the total PS pixels have more than the threshold STIP. This indicates that the CSLTM-ISS detected the most number of coherent PS pixels out of the three methods under comparison. The StaMPS method also detected a high percentage of coherent PS pixels, but the density of PS pixels is very low compared to CLSTM-ISS. Figure 10 shows a histogram of the frequency of PS pixels associated with the number of STIP. It is seen from the histogram that both StaMPS and CLSTM-SS have very low frequency of PS pixels (with STIP < 35). However, for CNN-ISS, the frequency is higher for STIP < 35 (shown in light brown color on the histogram). Table 6 shows the values associated with the above analysis. Table 7 shows an analysis of the number of common pixels detected by the three methods. It was observed that the StaMPS and CNN-ISS methods detected 46.74% pixels in common, whereas StaMPS and CLSTM-ISS detected 78.44% common pixels.

As a part of further analysis, the displacement estimates obtained using the three methods were compared. This would strengthen the understanding whether the proposed architectures can be used for generating displacement and/or velocity maps, which is the real purpose of PS selection. Figures 9 (d), (e) and (f) show the 1D velocity maps generated using StaMPS, CNN-ISS and CLSTM-ISS methods respectively. The velocity estimates obtained from the StaMPS method indicate subsidence in the Kathmandu city, which corresponds well with the pattern detected in Krishnan *et al.* (2018). The displacement pattern detected for the other part of the velocity map is also similar. However, for CNN-ISS, the velocity pattern is dissimilar from the pattern obtained using the StaMPS method, mostly at locations where non-PS pixels have been incorrectly detected as PS pixels (pixels with low STIP in Figure 9 (b)). The mismatch is marked using red color polygons in Figure 9 (e). Hence, the difference in estimated velocity may be attributed to the detection of some of the less coherent PS pixels, which further resulted in inaccurate phase unwrapping at some locations of the time series of interferograms, affecting velocity estimation. Moreover, for CLSTM-ISS, the velocity pattern is very similar to that detected using the StaMPS method. Additionally, due to the improved density of PS pixels, the interpretation of LOS velocity is improved.

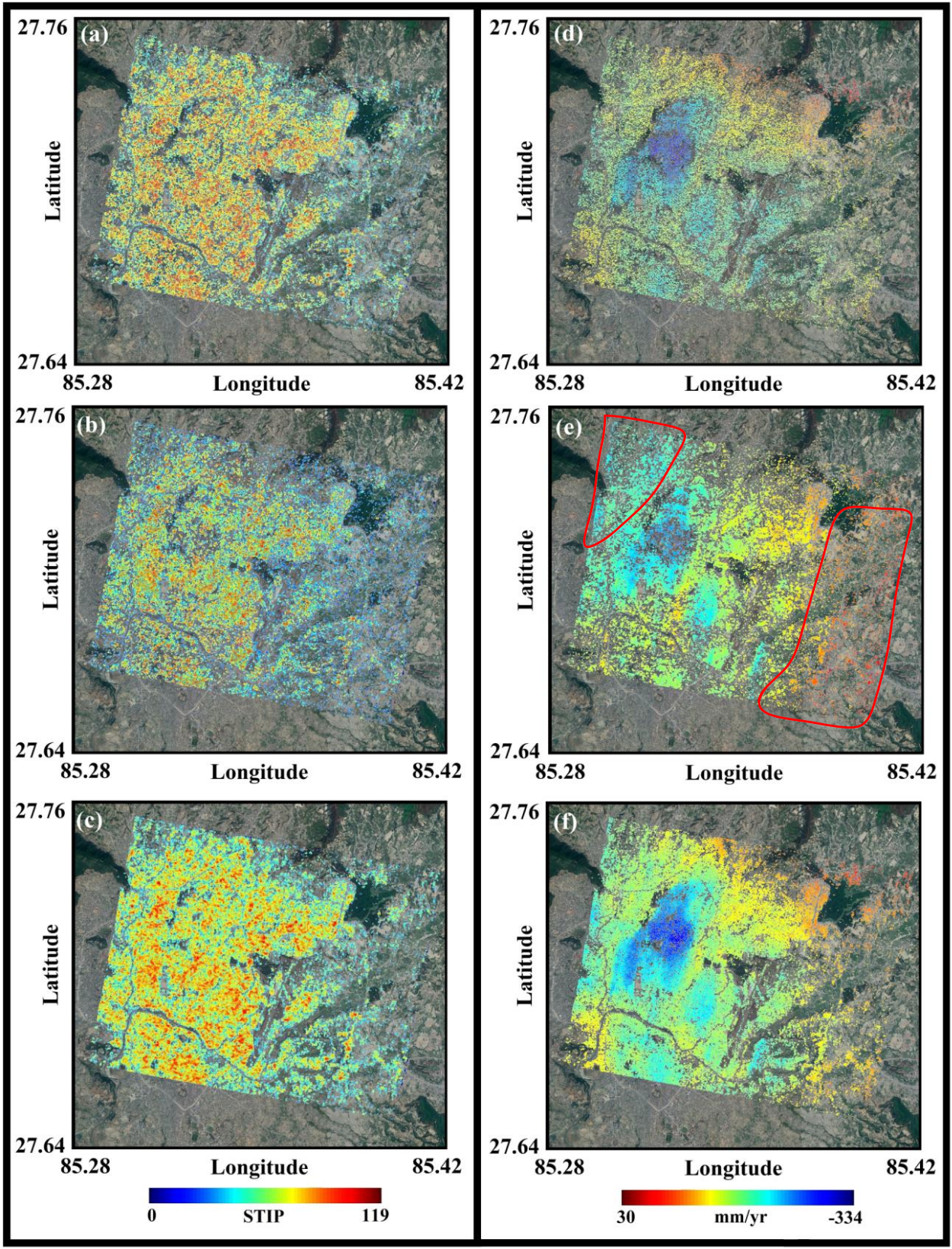

Figure 9: Number of STIP selected in the neighbourhood of PS pixels selected by (a) StaMPS, (b) CNN-ISS and (c) CLSTM-ISS. Estimated 1D LOS velocity of the PS pixels selected by (d) StaMPS, (e) CNN-ISS and (f) CLSTM-ISS.

Table 6. Number of STIP detected by StaMPS, CNN-ISS and CLSTM-ISS methods.

|                  | StaMPS | CNN-ISS | LSTM   | % of total PS |
|------------------|--------|---------|--------|---------------|
| Total PS detected | 38286  | 169241  | 192177 |               |
| PS with STIP >35 | 35413  | 137265  | 186435 |               |
| PS with STIP <35 | 2873   | 31976   | 5742   |               |

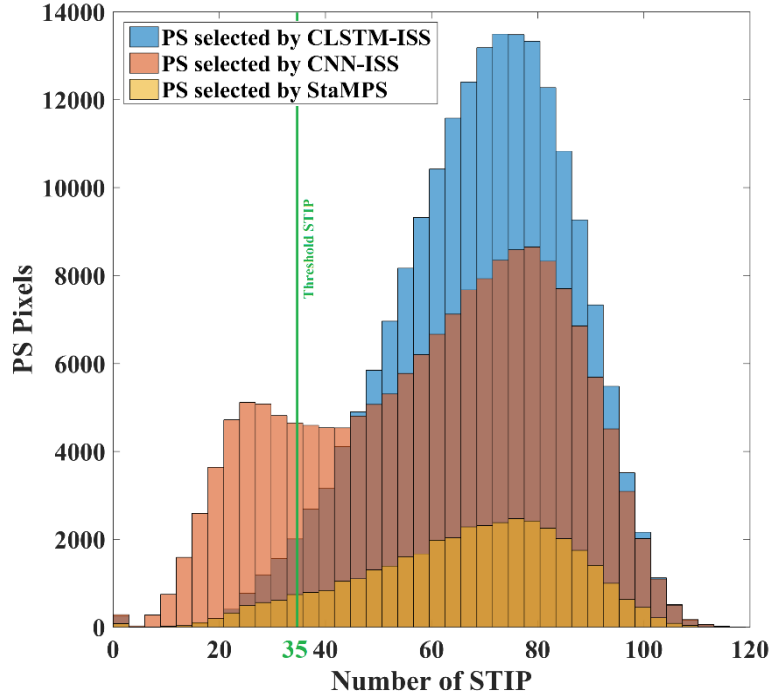

Figure 10. Histogram showing the frequency of the number of STIP detected by PS pixels for StaMPS, CNN-ISS and CLSTM-ISS methods. Green line marks the threshold STIP value.

Table 7. Common PS pixels detected between StaMPS, CNN-ISS and CLSTM-ISS methods. ∩ indicates common PS pixels in chosen methods.

| StaMPS ∩ CNN  | StaMPS ∩ LSTM |
|---------------|---------------|
| 18000 (46.74) | 30205 (78.44) |

Analysis of the time series displacement was further carried out to test the ability of the proposed architectures in estimating displacements at individual time steps. Figure 11 shows the time series plot of the Kathmandy city generated using StaMPS, CNN-ISS and CLSTM-ISS. The location marked as a red dot in Figure 6 (a) was selected for generating the time series displacement, having a temporal coverage of nearly eight months (24 March 2015 to 7 November 2015). This interval also included the occurrence of the Nepal earthquake on 25[th] April 2015). It was observed that the individual displacement estimates of the StaMPS and the CLSTM-ISS methods were similar for all the time steps.

A sudden change in the displacement magnitude is observed between the second and the third time step, which may well be due to the occurrence of the Nepal earthquake on 25th April 2015. However, the estimates generated from CNN-ISS were different for the second time step, and followed a similar displacement pattern after it. Moreover, the estimated displacement magnitude was different from that detected using the other two methods.

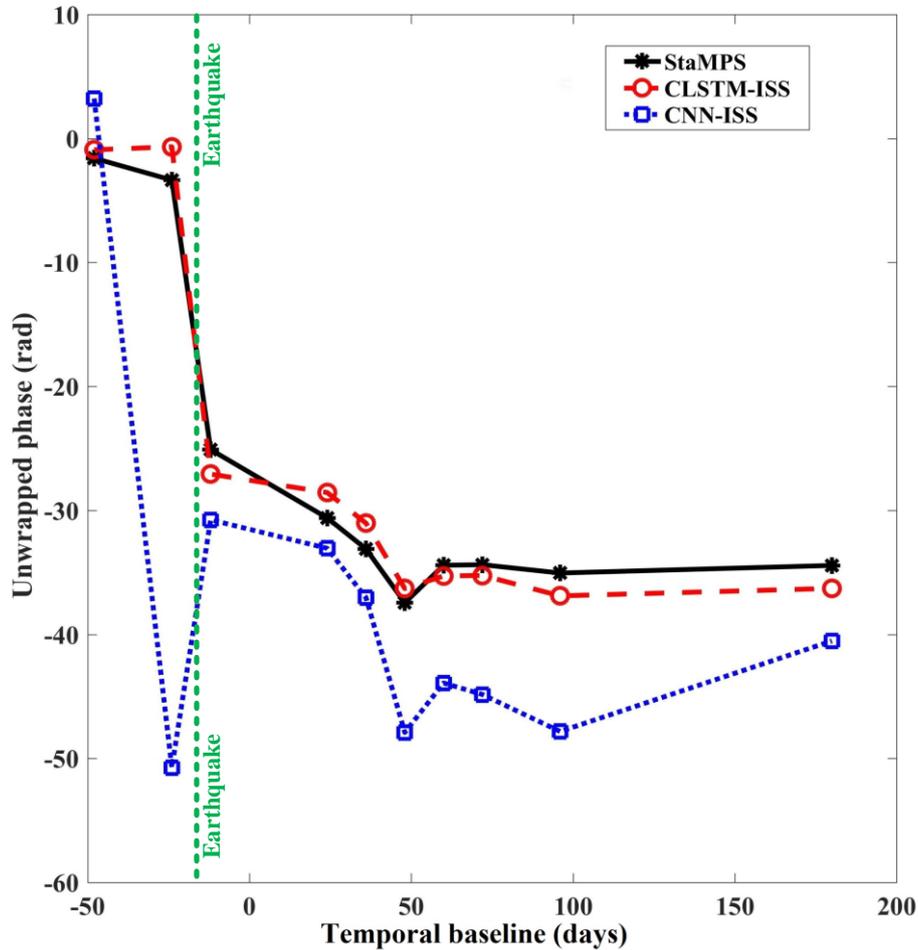

Figure 11. Time series displacement plot of the Kathmandu city generated using StaMPS, CNN-ISS and CLSTM-ISS. The location marked as a red dot in Figure 6 (a) is used for comparison. Temporal coverage of the generated time series is from 24 March 2015 to 7 November 2015. Green line marks the occurrence of the Nepal earthquake on 25th April 2015.

The above statistical analysis based on temporal coherence reveals that the CLSTM-ISS architecture detected the most number of reliable PS pixels among the three methods, and the percentage of such pixels is significantly high. Further, the analysis of the estimated 1D LOS velocities indicated that StaMPS and CLSTM-ISS methods detected similar velocity pattern, which conforms to the pattern detected during earlier studies. However, due to the increased density of reliable PS pixels, the displacement pattern is improved in case of CLSTM-ISS.

6. **Acknowledgement**

Authors are grateful to the ESA for providing Sentinel-1 IW images, and acknowledge the support received from the HPC facility at IIT Kanpur.

7. **Conclusion**

Two deep learning based architectures, namely CNN-ISS and CLSTM-ISS, were proposed in this study for PS pixel selection with a stack of differential interferograms as input. The architectures, learned using the labels obtained from StaMPS based PS selection, detected more number of PS pixels in urban areas containing man-made objects and under lengthening than that detected by the StaMPS method. The CLSTM-ISS approach detected PS pixels including almost all of the pixels detected as PS by the StaMPS method, indicating that nearly no true PS pixel identified by the StaMPS method was missed. The CSLSTM-ISS estimates, learnt using spatio-temporal behaviour of the interferometric phase, showed better separation of the two classes (PS and non-PS) than StaMPS and CNN-ISS. In this approach, there was an increase in the spatial density of PS pixels in case of cities covering man-made objects, and correct identification of PS pixels was observed even in difficult environments including the non-urban areas and the areas affected by foreshortening, layover and shadow. The above fact was also evident from the comparison of 1D velocity, time series displacement, noise statistics, and similarity of PS pixel detection. Apart from this, the computational efficiency of both the proposed methods was found to be better than all the benchmark methods used for PS selection. With the improved density and reliable estimation of the measurement pixels (PS pixels) along with the reduced computational time requirements, it is apparent that the CLSTM-ISS approach can provide better characterization of the deformation pattern and can support near real-time deformation monitoring applications.